\newif\ifprlout
\prlouttrue

\ifprlout
\documentclass[twocolumn,showpacs,superscriptaddress,preprintnumbers,prl,aps]{revtex4}
\else
\documentclass[12pt,twoside,showpacs,letterpaper,doublespace]{article}
\fi

\usepackage{amssymb,amsmath}
\usepackage{graphicx}
\usepackage{dcolumn}
\usepackage{bm}
\usepackage{lineno}
\usepackage{setspace}
\usepackage{overpic}

\def\B0    {\ensuremath{B^0}}

\def\Bs    {\ensuremath{B^0_s}}
\def\bBs  {\ensuremath{\bar{B}^0_s}}

\def\Jpsi{\ensuremath{J\!/\!\psi}}
\newcommand{\myto}{\kern -0.3em\to\kern -0.2em}
\def\BsJpsiPhi{\ensuremath{\Bs \myto \Jpsi \phi}}
\def\betas     {\ensuremath{\beta_s}}
\def\DGs       {\ensuremath{\Delta\Gamma_s}}
\newcommand{\Dms}{\ensuremath{\Delta m_s\ }}
\def\deltapara   {\ensuremath{\delta_{\parallel}}}
\def\deltaperp   {\ensuremath{\delta_{\perp}}}
\def\deltaS   {\ensuremath{\delta_{S}}}
\def\taus   {\ensuremath{\tau_{s}}}

\newcommand{\ifb}{\ensuremath{\mathrm{fb^{-1}}}}
\newcommand{\TeV}{\ensuremath{\matqq	hrm{Te\kern -0.1em V}}}
\newcommand{\TeVc}{\ensuremath{\mathrm{Te\kern -0.1em V\!/}c}}
\newcommand{\TeVcc}{\ensuremath{\mathrm{Te\kern -0.1em V\!/}c^2}}
\newcommand{\GeV}{\ensuremath{\mathrm{Ge\kern -0.1em V}}}
\newcommand{\GeVc}{\ensuremath{\mathrm{Ge\kern -0.1em V\!/}c}}
\newcommand{\GeVcc}{\ensuremath{\mathrm{Ge\kern -0.1em V\!/}c^2}}
\newcommand{\MeV}{\ensuremath{\mathrm{Me\kern -0.1em V}}}
\newcommand{\MeVc}{\ensuremath{\mathrm{Me\kern -0.1em V\!/}c}}
\newcommand{\MeVcc}{\ensuremath{\mathrm{Me\kern -0.1em V\!/}c^2}}

\newcommand{\stat}{\ensuremath{\text{(stat)}}}
\newcommand{\syst}{\ensuremath{\text{(syst)}}}

\newcommand{\mupmum}{\ensuremath{\mu^+\mu^-}}
\newcommand{\KpKm}{\ensuremath{K^+K^-}}

\newcommand{\CP}{\ensuremath{C\!P}}

\def\Babar {\mbox{\sl B\hspace{-0.4em} {\small\sl A}\hspace{-0.37em} \sl B\hspace{-0.4em} {\small\sl A\hspace{-0.02em}R}}}

\renewcommand{\epsilon}{\varepsilon}
\renewcommand{\theta}{\vartheta}

\begin{document}
\ifprlout
\else
\pagewiselinenumbers
\doublespace
\fi
\ifprlout
\title{
Measurement of the Bottom-Strange Meson Mixing Phase in the Full CDF Data Set} 
\affiliation{Institute of Physics, Academia Sinica, Taipei, Taiwan 11529, Republic of China}
\affiliation{Argonne National Laboratory, Argonne, Illinois 60439, USA}
\affiliation{University of Athens, 157 71 Athens, Greece}
\affiliation{Institut de Fisica d'Altes Energies, ICREA, Universitat Autonoma de Barcelona, E-08193, Bellaterra (Barcelona), Spain}
\affiliation{Baylor University, Waco, Texas 76798, USA}
\affiliation{Istituto Nazionale di Fisica Nucleare Bologna, $^{ee}$University of Bologna, I-40127 Bologna, Italy}
\affiliation{University of California, Davis, Davis, California 95616, USA}
\affiliation{University of California, Los Angeles, Los Angeles, California 90024, USA}
\affiliation{Instituto de Fisica de Cantabria, CSIC-University of Cantabria, 39005 Santander, Spain}
\affiliation{Carnegie Mellon University, Pittsburgh, Pennsylvania 15213, USA}
\affiliation{Enrico Fermi Institute, University of Chicago, Chicago, Illinois 60637, USA}
\affiliation{Comenius University, 842 48 Bratislava, Slovakia; Institute of Experimental Physics, 040 01 Kosice, Slovakia}
\affiliation{Joint Institute for Nuclear Research, RU-141980 Dubna, Russia}
\affiliation{Duke University, Durham, North Carolina 27708, USA}
\affiliation{Fermi National Accelerator Laboratory, Batavia, Illinois 60510, USA}
\affiliation{University of Florida, Gainesville, Florida 32611, USA}
\affiliation{Laboratori Nazionali di Frascati, Istituto Nazionale di Fisica Nucleare, I-00044 Frascati, Italy}
\affiliation{University of Geneva, CH-1211 Geneva 4, Switzerland}
\affiliation{Glasgow University, Glasgow G12 8QQ, United Kingdom}
\affiliation{Harvard University, Cambridge, Massachusetts 02138, USA}
\affiliation{Division of High Energy Physics, Department of Physics, University of Helsinki and Helsinki Institute of Physics, FIN-00014, Helsinki, Finland}
\affiliation{University of Illinois, Urbana, Illinois 61801, USA}
\affiliation{The Johns Hopkins University, Baltimore, Maryland 21218, USA}
\affiliation{Institut f\"{u}r Experimentelle Kernphysik, Karlsruhe Institute of Technology, D-76131 Karlsruhe, Germany}
\affiliation{Center for High Energy Physics: Kyungpook National University, Daegu 702-701, Korea; Seoul National University, Seoul 151-742, Korea; Sungkyunkwan University, Suwon 440-746, Korea; Korea Institute of Science and Technology Information, Daejeon 305-806, Korea; Chonnam National University, Gwangju 500-757, Korea; Chonbuk National University, Jeonju 561-756, Korea}
\affiliation{Ernest Orlando Lawrence Berkeley National Laboratory, Berkeley, California 94720, USA}
\affiliation{University of Liverpool, Liverpool L69 7ZE, United Kingdom}
\affiliation{University College London, London WC1E 6BT, United Kingdom}
\affiliation{Centro de Investigaciones Energeticas Medioambientales y Tecnologicas, E-28040 Madrid, Spain}
\affiliation{Massachusetts Institute of Technology, Cambridge, Massachusetts 02139, USA}
\affiliation{Institute of Particle Physics: McGill University, Montr\'{e}al, Qu\'{e}bec, Canada H3A~2T8; Simon Fraser University, Burnaby, British Columbia, Canada V5A~1S6; University of Toronto, Toronto, Ontario, Canada M5S~1A7; and TRIUMF, Vancouver, British Columbia, Canada V6T~2A3}
\affiliation{University of Michigan, Ann Arbor, Michigan 48109, USA}
\affiliation{Michigan State University, East Lansing, Michigan 48824, USA}
\affiliation{Institution for Theoretical and Experimental Physics, ITEP, Moscow 117259, Russia}
\affiliation{University of New Mexico, Albuquerque, New Mexico 87131, USA}
\affiliation{The Ohio State University, Columbus, Ohio 43210, USA}
\affiliation{Okayama University, Okayama 700-8530, Japan}
\affiliation{Osaka City University, Osaka 588, Japan}
\affiliation{University of Oxford, Oxford OX1 3RH, United Kingdom}
\affiliation{Istituto Nazionale di Fisica Nucleare, Sezione di Padova-Trento, $^{ff}$University of Padova, I-35131 Padova, Italy}
\affiliation{University of Pennsylvania, Philadelphia, Pennsylvania 19104, USA}
\affiliation{Istituto Nazionale di Fisica Nucleare Pisa, $^{gg}$University of Pisa, $^{hh}$University of Siena and $^{ii}$Scuola Normale Superiore, I-56127 Pisa, Italy}
\affiliation{University of Pittsburgh, Pittsburgh, Pennsylvania 15260, USA}
\affiliation{Purdue University, West Lafayette, Indiana 47907, USA}
\affiliation{University of Rochester, Rochester, New York 14627, USA}
\affiliation{The Rockefeller University, New York, New York 10065, USA}
\affiliation{Istituto Nazionale di Fisica Nucleare, Sezione di Roma 1, $^{jj}$Sapienza Universit\`{a} di Roma, I-00185 Roma, Italy}
\affiliation{Rutgers University, Piscataway, New Jersey 08855, USA}
\affiliation{Texas A\&M University, College Station, Texas 77843, USA}
\affiliation{Istituto Nazionale di Fisica Nucleare Trieste/Udine, I-34100 Trieste, $^{kk}$University of Udine, I-33100 Udine, Italy}
\affiliation{University of Tsukuba, Tsukuba, Ibaraki 305, Japan}
\affiliation{Tufts University, Medford, Massachusetts 02155, USA}
\affiliation{University of Virginia, Charlottesville, Virginia 22906, USA}
\affiliation{Waseda University, Tokyo 169, Japan}
\affiliation{Wayne State University, Detroit, Michigan 48201, USA}
\affiliation{University of Wisconsin, Madison, Wisconsin 53706, USA}
\affiliation{Yale University, New Haven, Connecticut 06520, USA}

\author{T.~Aaltonen}
\affiliation{Division of High Energy Physics, Department of Physics, University of Helsinki and Helsinki Institute of Physics, FIN-00014, Helsinki, Finland}
\author{B.~\'{A}lvarez~Gonz\'{a}lez$^z$}
\affiliation{Instituto de Fisica de Cantabria, CSIC-University of Cantabria, 39005 Santander, Spain}
\author{S.~Amerio}
\affiliation{Istituto Nazionale di Fisica Nucleare, Sezione di Padova-Trento, $^{ff}$University of Padova, I-35131 Padova, Italy}
\author{D.~Amidei}
\affiliation{University of Michigan, Ann Arbor, Michigan 48109, USA}
\author{A.~Anastassov$^x$}
\affiliation{Fermi National Accelerator Laboratory, Batavia, Illinois 60510, USA}
\author{A.~Annovi}
\affiliation{Laboratori Nazionali di Frascati, Istituto Nazionale di Fisica Nucleare, I-00044 Frascati, Italy}
\author{J.~Antos}
\affiliation{Comenius University, 842 48 Bratislava, Slovakia; Institute of Experimental Physics, 040 01 Kosice, Slovakia}
\author{G.~Apollinari}
\affiliation{Fermi National Accelerator Laboratory, Batavia, Illinois 60510, USA}
\author{J.A.~Appel}
\affiliation{Fermi National Accelerator Laboratory, Batavia, Illinois 60510, USA}
\author{T.~Arisawa}
\affiliation{Waseda University, Tokyo 169, Japan}
% ...

\collaboration{CDF Collaboration\footnote{With visitors from
$^a$Istituto Nazionale di Fisica Nucleare, Sezione di Cagliari, 09042 Monserrato (Cagliari), Italy,
$^b$University of CA Irvine, Irvine, CA 92697, USA,
$^c$University of CA Santa Barbara, Santa Barbara, CA 93106, USA,
$^d$University of CA Santa Cruz, Santa Cruz, CA 95064, USA,
$^e$Institute of Physics, Academy of Sciences of the Czech Republic, Czech Republic,
$^f$CERN, CH-1211 Geneva, Switzerland,
$^g$Cornell University, Ithaca, NY 14853, USA,
$^h$University of Cyprus, Nicosia CY-1678, Cyprus,
$^i$Office of Science, U.S. Department of Energy, Washington, DC 20585, USA,
$^j$University College Dublin, Dublin 4, Ireland,
$^k$ETH, 8092 Zurich, Switzerland,
$^l$University of Fukui, Fukui City, Fukui Prefecture, Japan 910-0017,
$^m$Universidad Iberoamericana, Mexico D.F., Mexico,
$^n$University of Iowa, Iowa City, IA 52242, USA,
$^o$Kinki University, Higashi-Osaka City, Japan 577-8502,
$^p$Kansas State University, Manhattan, KS 66506, USA,
$^q$Korea University, Seoul, 136-713, Korea,
$^r$University of Manchester, Manchester M13 9PL, United Kingdom,
$^s$Queen Mary, University of London, London, E1 4NS, United Kingdom,
$^t$University of Melbourne, Victoria 3010, Australia,
$^u$Muons, Inc., Batavia, IL 60510, USA,
$^v$Nagasaki Institute of Applied Science, Nagasaki, Japan,
$^w$National Research Nuclear University, Moscow, Russia,
$^x$Northwestern University, Evanston, IL 60208, USA,
$^y$University of Notre Dame, Notre Dame, IN 46556, USA,
$^z$Universidad de Oviedo, E-33007 Oviedo, Spain,
$^{aa}$CNRS-IN2P3, Paris, F-75205 France,
$^{bb}$Texas Tech University, Lubbock, TX 79609, USA,
$^{cc}$Universidad Tecnica Federico Santa Maria, 110v Valparaiso, Chile,
$^{dd}$Yarmouk University, Irbid 211-63, Jordan.
}}
\noaffiliation

\else
\begin{center}
\begin{Large}
Measurement of the Bottom-Strange Meson Mixing Phase in the Full CDF Data Set
\end{Large}
\end{center}
\vspace{1in}
\begin{center}
\begin{large}{The CDF Collaboration\\}
\end{large}
\today
\end{center}
\fi
\date{\today}
\begin{abstract}
We report a measurement of the bottom-strange meson mixing phase \betas\  using the time evolution of  $B^0_s \myto J/\psi(\myto\mu^+\mu^-) \phi (\myto K^+ K^-)$
decays in which the quark-flavor content of the bottom-strange meson is identified  at production. This measurement uses the full data set of proton-antiproton collisions at $\sqrt{s}= 1.96$ TeV collected by the Collider Detector experiment at the Fermilab Tevatron, corresponding to 9.6 fb$^{-1}$ of integrated luminosity.  We report confidence regions in the two-dimensional space of $\betas$ and the \Bs\ decay-width difference $\DGs$, and measure $\betas \in [-\pi/2, -1.51] \cup [-0.06,  0.30] \cup [1.26, \pi/2]$ at the 68\% confidence level, in agreement with the standard model expectation. Assuming the standard model value of \betas, we also determine $\DGs =  0.068 \pm 0.026\stat \pm0.009\syst$ ps$^{-1}$ and the mean \Bs\ lifetime,  $\tau_s = 1.528 \pm 0.019\stat \pm0.009\syst$ ps, which are consistent and competitive with determinations by other experiments.
\end{abstract}

\ifprlout
\maketitle
\fi
The noninvariance of the physics laws under the simultaneous transformations of parity and charge conjugation (\CP\ violation) is accommodated in the standard model (SM) through the presence of a single irreducible complex phase in the weak-interaction couplings of quarks. A broad class of generic extensions of the SM is expected to naturally introduce additional sources of \CP\ violation that should be observable, making \CP--violation studies promising to search for experimental indications of new particles or interactions. Thus far, \CP\ violation has been established in transitions of strange and bottom hadrons, with effects consistent with the SM interpretation~\cite{pdg,hfag,Antonelli:2009ws}. Much less information is available for bottom-strange mesons, $B^0_s$. Studies of $B^0_s$--$\overline{B}^0_s$ flavor oscillations are unique in that they probe the quark-mixing (Cabibbo-Kobayashi-Maskawa, CKM) matrix element $V_{ts}$, which directly enters the mixing amplitude. Large non-SM enhancements of the mixing amplitude are excluded by the precise determination of the oscillation frequency in 2006~\cite{Abulencia:2006ze}. However, non-SM particles or couplings involved in the mixing may also increase the size of the observed \CP\ violation by enhancing the mixing phase $\beta_s = \arg[-(V_{ts}V_{tb}^*)/(V_{cs}V_{cb}^*)]$~\cite{phis} with respect to the value expected from the CKM hierarchy, $\beta_s^{\rm{SM}} \approx  0.02$~\cite{hfag}, henceforth referred to as `SM expectation'. A non-SM enhancement of \betas\  would also decrease 
the size of the decay-width difference between the light and heavy mass eigenstates of the \Bs\ meson, $\DGs = \Gamma_L - \Gamma_H$. The values of the mixing phase and width difference are loosely constrained, and currently the subject of intense experimental activity. The analysis of the time evolution of \BsJpsiPhi\ decays provides the most effective determination of \betas\ and \DGs~\cite{jpsiphi}.  Assuming negligible contributions from sub-leading decay amplitudes~\cite{Faller:2008gt}, the underlying $b \myto c\bar{c}s$ quark transition is 
dominated by a single real amplitude, making \betas\ the sole \CP-violating phase observable, through the interference between the amplitudes of decays occurring with and without oscillations.\par
The first determinations of \betas, by the CDF and D0 experiments, suggested a mild deviation from the SM expectation~\cite{sin2betas-early}. The interest in this measurement increased further recently, because of  the 3.9$\sigma$ departure from the SM expectation of the dimuon asymmetry observed by D0 in semileptonic decays of $B^0_{(s)}$ mesons~\cite{asl}, which is tightly correlated with \betas, if generated in the \Bs\ sector~\cite{phis}. While updated measurements in \BsJpsiPhi\ decays~\cite{CDF:2011af,Abazov:2011ry,LHCb:2011aa, ATLAS} showed increased consistency with the SM, more precise experimental information is needed for a conclusive interpretation. \par
In this Letter we report a measurement of \betas;  \DGs; the mean lifetime of  heavy and light \Bs\ mass eigenstates, $\tau_s = 2/(\Gamma_H+\Gamma_L)$; and the angular momentum composition of the signal sample using the final data set collected by the CDF experiment at the Tevatron proton-antiproton collider, corresponding to an integrated luminosity of 9.6 \ifb. The analysis closely follows a previous measurement in a subset of the present data \cite{CDF:2011af}, and introduces an improved determination of the sample composition based on a new study of the $K^+K^-$ and $J/\psi K^+K^-$ mass distributions.\par
The CDF II detector is a magnetic spectrometer  surrounded by electromagnetic and hadronic calorimeters and muon detectors that has cylindrical geometry with
forward-backward symmetry. Charged particle trajectories (tracks) are reconstructed 
using single- and double-sided silicon microstrip sensors arranged in seven cylindrical layers~\cite{silicon} and an open cell drift chamber with 96 layers of sense wires~\cite{COT},  all immersed in a 1.4~T axial magnetic field.
The resolution on the momentum component transverse to the beam, $p_T$, is $\sigma_{p_{T}}/p^2_{T} \approx 0.07\%$ ($p_T$ in GeV/$c$), corresponding to a mass resolution of  our $B^0_s$ signal of about 9 MeV/$c^2$. Muons with $p_T>1.5$ GeV/$c$ are detected in multiwire drift chambers~\cite{muoni}.
A time-of-flight detector identifies  charged particles with $p_T< 2$ GeV/$c$~\cite{tof}, complemented by the ionization-energy-loss measurement in the drift chamber at higher transverse momenta. The combined identification performance corresponds to a separation between charged kaons and pions of approximately two Gaussian standard deviations, nearly constant in the relevant momentum range. Events enriched in $\Jpsi \myto \mupmum$ decays  are recorded using a low-$p_T$ dimuon online selection (trigger) that requires two oppositely-charged particles reconstructed in the drift chamber matched to muon chamber track segments, with a dimuon mass between 2.7 and 4.0~\GeVcc.\par
In the analysis, two tracks matched to muon pairs are required to be consistent with a $\Jpsi \myto \mu^+\mu^-$ decay, with dimuon mass $3.04< m_{\mu\mu}< 3.14$ \GeVcc. These are combined with another pair of tracks consistent with a $\phi \myto \KpKm$ decay, $1.009< m_{KK}< 1.028$ \GeVcc, in a kinematic fit to a common vertex. A dimuon mass constraint to the known \Jpsi\ mass~\cite{pdg} improves the \Bs\  mass  resolution. 
An artificial neural network (NN) classifier~\cite{CDF:2011af} combines multiple discriminating variables into a single quantity that statistically separates
the signal from the dominant background from combinations of real $\Jpsi$ decays with random track pairs and a minor component of random four-track combinations (both collectively referred to as combinatorics). 
 The NN is trained with simulated events for the signal and data from sidebands in \Bs\ mass,  $[5.29,5.31] \cup [5.42,5.45]$ GeV/$c^2$, 
for the background.  In decreasing order of discriminating power, the input variables to the NN  include kinematic quantities, muon and hadron particle identification information, and vertex fit quality parameters. \par Figure~\ref{fig:mass} shows the $J/\psi K^+ K^-$ mass distribution from the final sample of candidates that pass an NN threshold chosen as to maximize the sensitivity to the measurement of \betas~\cite{CDF:2011af}. The distribution shows a signal of approximately 11~000 decays, above a fairly constant background dominated by the prompt combinatorial component, and smaller contributions from mis-reconstructed $B$ decays.
\begin{figure}[htbp]
\centerline{
  \makebox{
     \includegraphics[width=0.5\textwidth]{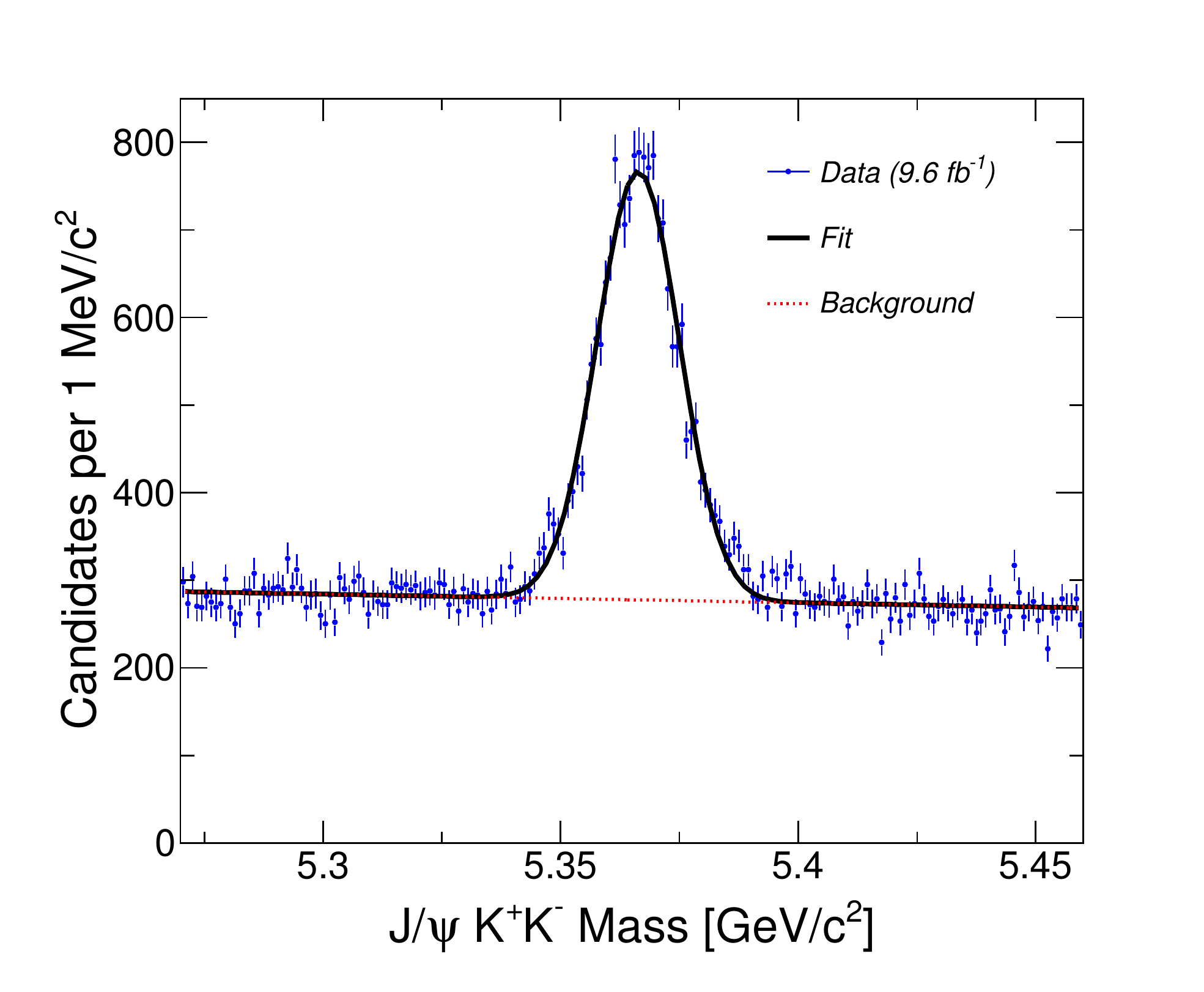}
   }
} 
\caption{\label{fig:mass} (Color online) Distribution of $\Jpsi K^+K^-$ mass with fit projection
overlaid. } 
\end{figure}
\par We determine the quantities of interest using a fit to the time evolution of bottom-strange mesons. The differences in time evolution of states initially produced as a  $B^0_s$ or $\overline{B}^0_s$ meson are included in the fit as well as the differences between decays that result in a \CP-odd or \CP-even combination of the $\Jpsi \phi$ angular momenta. 
The proper decay time of a \Bs\ candidate is a fit observable calculated as $t = M L_{xy}/p_T$, 
where $L_{xy}$ is the distance from the primary vertex to the \Bs\ decay vertex, projected onto the \Bs\ momentum in the plane transverse to the beam, $\vec{p}_T$; and $M$ is the known mass of the $\Bs$~meson~\cite{pdg}. The proper decay-time uncertainty, $\sigma_t$, is calculated from the measurement uncertainties in  $L_{xy}$. Because the $B^0_s$ meson has spin zero and \Jpsi\ and $\phi$ have spin one,  the \BsJpsiPhi\ decay involves three possible angular momentum states of the $\Jpsi \phi$ system. These are combined into three polarization amplitudes,  longitudinal polarization ($A_0$),  and transverse polarization with the linear polarization vectors of the
vector mesons parallel ($A_\parallel$)  or perpendicular ($A_\perp$) to each other. The first two states are \CP\ even, while the last state is \CP\ odd. A \CP-odd state can also be produced by a nonresonant $\KpKm$ pair or can originate from the decay of the spin-0 $f_0(980)$ meson, which results in another independent decay amplitude, the $S$-wave $A_S$. \par
To enhance the sensitivity to \betas, the time-evolution of the four decay amplitudes along with six interference terms
is fitted simultaneously by exploiting differences in the distribution of the kaons' and muons' decay angles. The angles are parametrized in the transversity basis, $\vec{\rho}=(\cos\Theta,\Phi,\cos\Psi)$~\cite{transversity}, which allows a convenient separation of the \CP-even and \CP-odd terms in the likelihood.  Reference~\cite{Azfar:2010nz} details the expression for the decay rate differential in the decay time and angles. The rate is a function of the physics parameters of interest, \betas, \DGs, \taus, and the decay amplitudes with their \CP-conserving phases. For these we choose $A_0$ to be real and define the \CP-conserving phases as $\deltapara=\arg{(A_\parallel\!/\!A_0)}$, $\deltaperp=\arg{(A_\perp\!/\!A_0)}$  and $\deltaS=\arg{(A_S\!/\!A_0)}$. The decay rate is also a function of the \Bs\ mixing frequency, which is a fit parameter constrained to the experimental value measured by CDF, $\Dms = 17.77 \pm 0.12$~ps$^{-1}$~\cite{Abulencia:2006ze}.\par 
The flavor of the meson at the time of production is inferred by two independent classes of flavor tagging algorithms~\cite{CDF:2011af}, which exploit
specific features of the incoherent  production of $b\bar{b}$ quarks-pairs in $p\bar{p}$ collisions. 
Using flavor conservation of the strong interaction, the opposite-side flavor tag (OST) infers the signal production flavor from the decay products of the $b$ hadron produced by the other $b$~quark in the event by using the charge of muons or
electrons from semileptonic $B$~decays or the net charge of the opposite-side jet. 
The same-side kaon tag (SSKT) deduces the signal production flavor by exploiting charge-flavor correlations of the neighboring kaons produced during its fragmentation.  The fraction of candidates tagged by a combination of OST algorithms totals $\epsilon_{\rm{OST}} = (92.8\pm0.1)\%$.
The probability of wrongly-tagging the meson, $w_{\rm{OST}}$,  is determined per event and calibrated using 82~000 $B^\pm \myto \Jpsi(\myto \mu^+\mu^-) K^\pm$ decays fully reconstructed in the same  sample as the signal~\cite{Sabato}. Because the $B^{\pm}$ does not oscillate, the OST tag is compared with the actual flavor, 
known from the charge of the $K^{\pm}$ meson. A single scale factor  that matches the predicted mistag probability to the one observed in data is then determined to be $1.085 \pm 0.035$. The observed averaged dilution, ${\cal D}_{\rm{OST}} = 1 -2 w_{\rm{OST}}$, equals $(12.3 \pm 0.6)\%$ including the scale factor, resulting in a tagging power of $\epsilon_{\rm{OST}}{\cal D}^2_{\rm{OST}} = (1.39 \pm 0.05)\%$. The SSKT algorithms tag a smaller fraction of candidates, $\epsilon_{\rm{SSKT}} = (52.2 \pm 0.7)\%$,  with better precision. 
In the  \BsJpsiPhi\ sample an average dilution of ${\cal D}_{\rm{SSKT}} = (25.9 \pm 5.4)\%$  is achieved including 
a $0.94 \pm 0.20$ scale factor obtained by measuring the $B^0_s$ oscillation amplitude in approximately 11~000 (1~850)  $\Bs \myto D_s^- \pi^+ (\pi^+\pi^-)$  decays reconstructed in the data corresponding to the first 5.2 fb$^{-1}$~\cite{CDF:2011af}. The resulting SSKT tagging power is $\epsilon_{\rm{SSKT}}{\cal D}^2_{\rm{SSKT}}=(3.5 \pm 1.4)\%$.
Higher instantaneous luminosity conditions in later data resulted in a reduced trigger efficiency for hadronic $B^0_s$ decays. Hence, the additional sample of $\Bs \myto D_s^- \pi^+(\pi^+\pi^-)$ decays is too limited for a significant test of the SSKT performance.  Because the SSKT calibration is known for early data only, we conservatively restrict its use to the events collected in that period. Simulation shows that this results in a degradation in \betas\  resolution not exceeding 15\%.\par 
The unbinned maximum likelihood fit uses 9 observables from each event to determine 32 parameters including $\beta_s$ and $\Delta\Gamma$,  other physics parameters such as $\Bs$ lifetime, amplitudes and phases,  and several other quantities, called \emph{nuisance} parameters, such as tagging dilution scale factors. The fit uses the information of the reconstructed $\Bs$ candidate mass and its uncertainty, $m$ and $\sigma_m$;
the $\Bs$ candidate proper decay time and its uncertainty, $t$ and $\sigma_{t}$; the three transversity angles, $\vec{\rho}$;
and tag information, ${\cal D}$ and $\xi$; where
${\cal D}$ is the event-specific dilution given by
the mistag probability, and $\xi$ is the
tag decision. Both tagged and untagged events are used
in the fit.  The single-event likelihood is described in terms of signal, $P_s$,  and background, $P_b$, 
probability density functions (density henceforth) as
 \begin{eqnarray}
        \mathcal{L} &&\propto  f_s P_s(m|\sigma_m) P_s(t,\vec\rho,\xi|{\cal D}, \sigma_{t}) 
                P_s(\sigma_{t}) P_s({\cal D}) \nonumber \\
           &&+(1-f_s) P_b(m) P_b(t| \sigma_{t}) P_b(\vec\rho) 
                P_b(\sigma_{t}) P_b({\cal D}),
 \label{eqn:fullPDF}
\end{eqnarray}
where $f_s$ is the fraction of signal events. 
The signal mass density, $P_s(m|\sigma_m)$, is parametrized as a single
Gaussian with a width determined independently for each
candidate. The background mass density, $P_b(m)$, is parametrized
as a straight line.  The time and angular dependence of the signal,
$P_s(t,\vec\rho,\xi,|{\cal D},\sigma_{t})$, for a single flavor tag
are written in terms of two densities, $P$ for $\Bs$ and $\bar{P}$ for
$\bBs$, as
\begin{equation}
  \Bigg( \frac{1+\xi{\cal D}}{2} P(t,\vec\rho| \sigma_{t})   
 +\frac{1-\xi{\cal D}}{2}\bar{P}(t, \vec\rho| \sigma_{t})\Bigg) \epsilon(\vec\rho),
\label{eqn:signalPDF}
\end{equation}
which is extended to the case of OST and SSKT independent flavor tags.  Acceptance effects on the transversity
angle distributions are modeled with an empirical three-dimensional joint probability density function
extracted from simulation, $\epsilon(\vec\rho)$, whose largest excursions do not exceed 15\%. The time and angular distributions for flavor-tagged \Bs\ (\bBs) decays,  $P$ ($\bar{P}$),  are given by 
 the normalized decay rate as functions of decay time and 
 transversity angles of Ref.~\cite{Azfar:2010nz}, assuming no direct $\textit{CP}$~violation in the decay.  
  Building on previous measurements~\cite{Aaltonen:2010pj}, we model the decay-time density for the background, $P_b(t| \sigma_{t})$,
with a $\delta$-function at $t=0$, one positive, and two negative exponential functions.
All time-dependent terms are convolved with a proper time resolution function, 
modeled as a sum of two Gaussians with common mean and independent widths determined by the fit.
The resulting decay-time resolution 
is equivalent to that of a Gaussian distribution with 90 fs standard deviation. 
 The background angular probability density is determined from \Bs\ mass sideband events to factorize as
$P_b(\vec\rho) = P_b(\cos\Theta)P_b(\Phi)P_b(\cos\Psi)$.
The distributions of the decay-time uncertainty and the event-specific dilution 
differ for signal and background events, thus their densities are
explicitly included in the likelihood.  The probability density functions of the decay-time uncertainties,
$P_s(\sigma_{t})$ and $P_b(\sigma_{t})$, are described with an empirical model from an independent fit to the data. 
The signal density, $P_s({\cal D})$, is determined from binned background-subtracted signal distributions, while the background density, 
$P_b({\cal D})$, is modeled from candidates in the signal sidebands.
Potential sources of systematic uncertainties, associated with imprecisely known calibration factors of tagging dilutions, are taken into account by floating these factors in the fit within Gaussian constraints.\par
The likelihood function shows two equivalent global maxima, corresponding to the solutions with positive and negative value of \DGs,  and additional local maxima generated by approximate symmetries~\cite{Azfar:2010nz}. Multiple solutions make  the estimation of parameters and their uncertainties challenging with limited sample size. 
If \betas\ is fixed to its SM value, the fit shows unbiased estimates and Gaussian uncertainties for \DGs,
\taus, polarization amplitudes, and the phase \deltaperp, yielding \begin{align*}
\taus &= 1.528 \pm 0.019\stat \pm 0.009\syst\, \rm{ps,}\\
\DGs &= 0.068 \pm 0.026\stat \pm 0.009\syst\, \rm{ps}^{-1},\\
|A_0|^2 &= 0.512 \pm 0.012\stat \pm 0.018\syst,\\
|A_\parallel|^2 &= 0.229 \pm 0.010\stat \pm 0.014\syst,\\
\deltaperp &= 2.79 \pm 0.53\stat \pm 0.15\syst.
\end{align*}
The correlation between $\tau_s$ and $\DGs$ is 0.52.
We do not report a measurement of \deltapara. The fit determines $\deltapara \approx \pi$,  but the estimate is biased and its uncertainty is non-Gaussian because the likelihood symmetry under the $\deltapara \myto 2\pi - \deltapara$ transformation~\cite{Azfar:2010nz} results in multiple maxima in the vicinity of $\deltapara = \pi$.  Systematic uncertainties include mismodeling of the signal mass model, decay-time resolution,  acceptance description, and angular distribution of the background;
an 8\% contamination by $\B0 \myto \Jpsi K^*(892)^0$ and  $B^0 \myto \Jpsi K^+\pi^-$ decays misreconstructed as \BsJpsiPhi\ decays; and silicon detector misalignment.
For each source, uncertainties are determined by comparing the fit results from simulated samples in which the systematic effect is introduced in the model and samples simulated according to the default model. The uncertainty on the \DGs\ measurement is dominated by the mismodeling of the background decay time. The largest contribution to the uncertainty on \taus\ is the effect of  
silicon detector misalignment. The angular acceptance model dominates the systematic uncertainties on the amplitudes.\par
If \betas\ is free to float in the fit, tests in statistical trials show that the maximum likelihood estimate 
is biased for the parameters of interest, and the biases depend on the true values 
of the parameters. Hence, we determine confidence regions in the \betas\ and $(\betas, \DGs)$ spaces by using a profile-likelihood ratio statistic as a $\chi^2$ variable and considering all other likelihood variables as nuisance parameters. The profile-likelihood ratio distributions observed in simulations deviate
from the expected $\chi^2$ distribution,  yielding confidence regions that contain the true values of the parameters 
with lower probability than the nominal confidence level. In addition, the profile-likelihood ratio distribution depends on the true values 
of the unknown nuisance parameters.  We use a large number of statistical trials to derive the profile-likelihood ratio distribution of our data. 
The effect of nuisance parameters is accounted for by randomly sampling their 30-dimensional space within
$5\sigma$ of their estimates in data and using the most conservative of the resulting profile-likelihood ratio distributions to derive the final confidence regions. This procedure ensures that the confidence regions have nominal statistical coverage whatever the configuration of nuisance parameters values and increases the size of the $\betas$ confidence interval by about 40\%.
We determine the confidence level for  $32\times48$ evenly spaced points in $\betas\in[-\pi/2, \pi/2]$ and
$\DGs\in[-0.3,0.3]$ ps$^{-1}$ and smoothly interpolate between them to obtain a continuous region (Fig.~\ref{fig:contour}).
Assuming the standard model values for \betas\ and \DGs, the probability to observe a profile-likelihood ratio equal to or higher than observed in data is 54\%. By treating \DGs\ as a nuisance parameter, we also obtain $\beta_s\in[-\pi/2,-1.51]\cup[-0.06,0.30]\cup [1.26,\pi/2]$ at the 68\% C.L., and $\betas \in [-\pi/2, -1.36] \cup [-0.21,0.53] \cup [1.04,\pi/2]$ at the 95\% C.L.
\begin{figure}
\begin{overpic}[width=0.5\textwidth]
{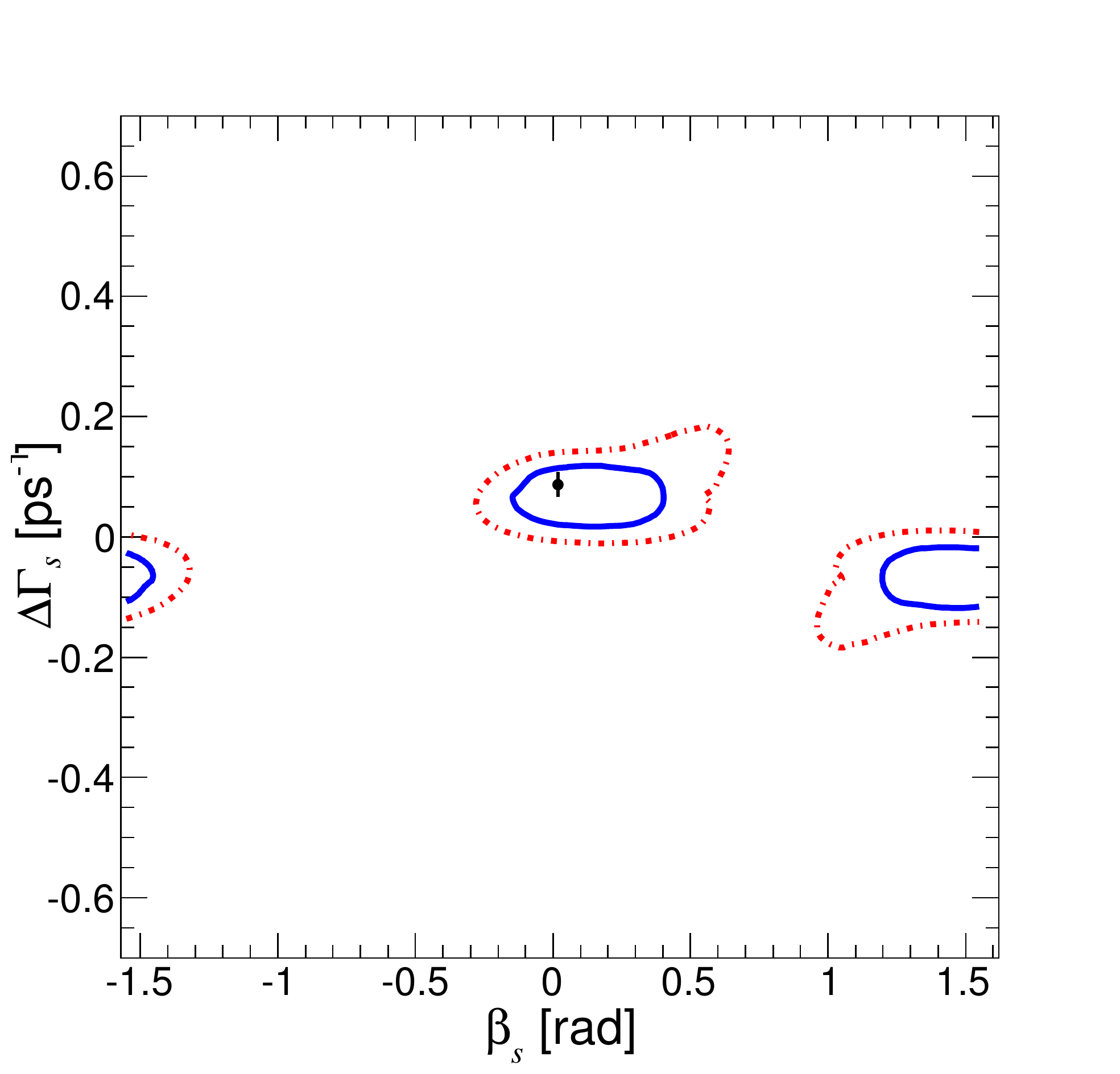}
\put(16,12){\includegraphics[width=0.19\textwidth]%
{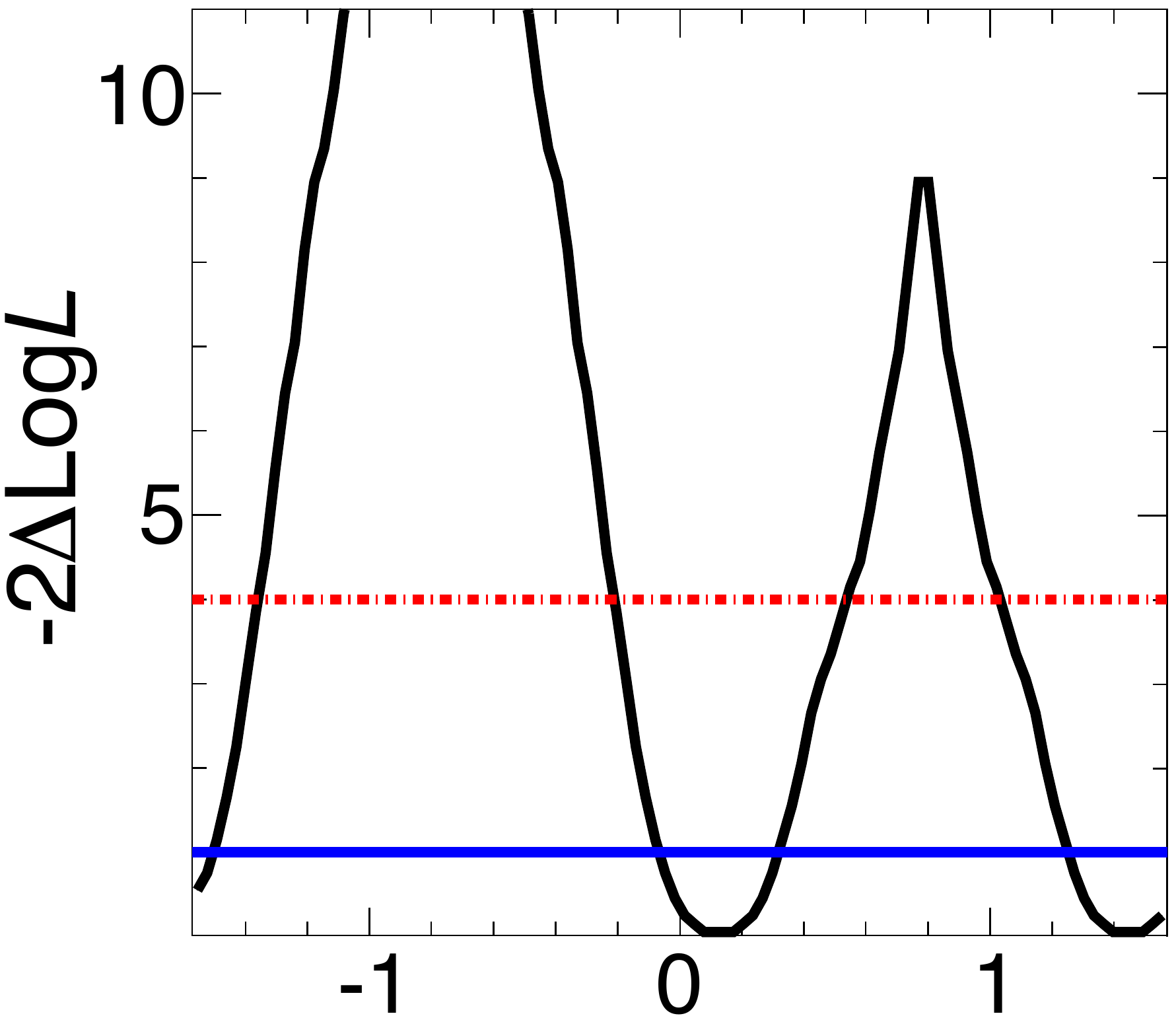}}
\end{overpic}
\caption{\label{fig:contour}
(Color online) Confidence regions at the 68\% (solid) and 95\% C.L.(dashed)  in the (\betas, \DGs) plane (main panel). The standard model prediction is  shown as a circle with error bars~\cite{LenzNierste}. The inset shows the coverage-corrected profile-likelihood ratio as a function of $\betas$, in which \DGs\ is treated as all other nuisance parameters.}
\end{figure}
The fraction of $S$-wave in the \KpKm\ mass range 1.009--1.028~\GeVcc\ 
is determined from the angular information to be consistent with zero with $\mathcal{O}(2\%)$ uncertainty, which is in agreement with our previous determination~\cite{CDF:2011af} and the LHCb and ATLAS results~\cite{LHCb:2011aa, ATLAS}, and inconsistent with the D0 determination~\cite{Abazov:2011ry}. An auxiliary simultaneous fit of the $K^+K^-$ and $\Jpsi K^+K^-$ mass distributions~\cite{Mirco} that includes the full resonance structure of the $B^0 \myto \Jpsi K^+\pi^-$ decay~\cite{Aubert:2004cp} is performed. The $K^+K^-$ mass is fit in a range enlarged to 0.988--1.2 GeV/$c^2$ using a relativistic Breit-Wigner distribution for the $\phi$ meson, the shape suggested in Ref.~\cite{flatte} for the $f_0(980)$ meson, and an empiric shape determined from data for the combinatorial background.
In the 1.009--1.028 GeV/$c^2$ mass range, this fit determines a $(0.8 \pm 0.2(\rm{stat}))\%$ $K^+K^-$ $S$-wave contribution in agreement with the central fit, and a contamination from mis-identified $B^0$ decays of  $(8.0 \pm 0.2(\rm{stat}))\%$, which is significantly larger than the 1--2\% values typically derived assuming only $P$-wave $B^0$ decays~\cite{CDF:2011af, Abazov:2011ry}. If neglected,  this additional $B^0$ component could mimic a larger $K^+K^-$ $S$-wave than present.\par
 In summary we report the final CDF results on the \Bs\ mixing phase and
decay width difference from the time-evolution of flavor-tagged $\BsJpsiPhi$ decays reconstructed in the full Tevatron Run II data set.
This analysis improves  and supersedes the previous CDF measurement obtained in a subset of the present data \cite{CDF:2011af}.
Considering \DGs\ as a nuisance parameter, and using the recent determination of the sign of \DGs~\cite{Aaij:2012eq}, we find  $-0.06< \beta_s< 0.30$ at the 68\% C.L.
Assuming a SM value for \betas, we also report precise measurements of decay-width difference,  $\DGs =  0.068 \pm 0.026\stat \pm0.009\syst$ ps$^{-1}$, and 
mean \Bs\ lifetime,  $\tau_s = 1.528 \pm 0.019\stat \pm0.009\syst$ ps. 
All results are consistent with expectations and with determinations of the same quantities from other experiments~\cite{Abazov:2011ry,LHCb:2011aa, ATLAS},  and significantly improve the knowledge of the phenomenology on \CP\ violation in \Bs\ mixing.\par

We thank the Fermilab staff and the technical staffs of the participating institutions for their vital contributions. This work was supported by the U.S. Department of Energy and National Science Foundation; the Italian Istituto Nazionale di Fisica Nucleare; the Ministry of Education, Culture, Sports, Science and Technology of Japan; the Natural Sciences and Engineering Research Council of Canada; the National Science Council of the Republic of China; the Swiss National Science Foundation; the A.P. Sloan Foundation; the Bundesministerium f\"ur Bildung und Forschung, Germany; the Korean World Class University Program, the National Research Foundation of Korea; the Science and Technology Facilities Council and the Royal Society, UK; the Russian Foundation for Basic Research; the Ministerio de Ciencia e Innovaci\'{o}n, and Programa Consolider-Ingenio 2010, Spain; the Slovak R\&D Agency; the Academy of Finland; and the Australian Research Council (ARC). 



\begin{thebibliography}{99}
\bibitem{phis}
Rigorously, the mixing phase is the phase of the off-diagonal element of the mixing transition matrix $M_{12}$, which approximates $2\beta_s$ within $\mathcal{O}(10^{-3})$ corrections. Another often used quantity is $\phi_s \equiv \arg[-M_{12}/\Gamma_{12}]$, where  $\Gamma_{12}$ is the decay width of $B^0_s$ and $\overline{B}^0_s$ mesons into common final states,  which governs the asymmetry in $B^0_s$ semileptonic decays of $B^0_s$ mesons. If significant non-SM contributions affect the mixing amplitude, the relation $\phi_s \approx -2 \beta_s$ holds between observed quantities.
\end{thebibliography}


\end{document}